\documentclass[twoside,12pt]{article}
\usepackage{indentfirst}
\usepackage{bm}
\usepackage{graphicx}
\usepackage{amsmath}
\usepackage{amssymb}

\begin{document}


\thispagestyle{empty} \vspace*{0.8cm}\hbox
to\textwidth{\vbox{\hfill\huge\sf Chinese Physics B\hfill}}
\par\noindent\rule[3mm]{\textwidth}{0.2pt}\hspace*{-\textwidth}\noindent
\rule[2.5mm]{\textwidth}{0.2pt}

\begin{center}
\LARGE\bf Statistical physics of
hard combinatorial optimization: The vertex cover problem
\end{center}

\footnotetext{\hspace*{-.45cm}\footnotesize $^*$This work was supported  by the National Basic Research Program
of China (No. 2013CB932804), the Knowledge Innovation Program of
Chinese Academy of Sciences (No.~KJCX2-EW-J02),
and the National Science Foundation of China
(grant Nos.~11121403, 11225526).}
\footnotetext{\hspace*{-.45cm}\footnotesize $^\dag$
 Corresponding author. E-mail: {\tt zhouhj@itp.ac.cn}}

\begin{center}
  \rm Jin-Hua Zhao and Hai-Jun Zhou$^\dagger$
\end{center}

\begin{center}
\begin{footnotesize} \sl
State Key Laboratory of Theoretical Physics,
Institute of Theoretical Physics,
Chinese Academy of Sciences,
Beijing 100190, China
\end{footnotesize}
\end{center}


\vspace*{2mm}

\begin{center}
\begin{minipage}{15.5cm}
\parindent 20pt\footnotesize
Typical-case computation complexity is a research topic at
the boundary of computer science, applied mathematics, and
statistical physics. In the last twenty years the
replica-symmetry-breaking mean field theory of spin glasses
and the associated message-passing algorithms have greatly
deepened our understanding of typical-case computation
complexity. In this paper we use the vertex cover problem, a
basic nondeterministic-polynomial (NP)-complete combinatorial
optimization problem of wide application, as an example to
introduce the statistical physical methods and
algorithms. We do not go into the technical details
but emphasize mainly the intuitive physical meanings of the
message-passing equations. A nonfamiliar reader shall be
able to understand to a large extent the physics behind the
mean field approaches and to adjust them in solving other
optimization problems.
\end{minipage}
\end{center}

\begin{center}
\begin{minipage}{15.5cm}
\begin{minipage}[t]{2.3cm}{\bf Keywords:}\end{minipage}
\begin{minipage}[t]{13.1cm}
spin glass, energy minimization, replica symmetry breaking,
belief propagation, survey propagation
\end{minipage}\par\vglue8pt
{\bf PACS:}
89.20.Ff, 75.10.Nr, 02.10.Ox, 05.10.-a
\end{minipage}
\end{center}

\clearpage

\section{Introduction}


The notion of computation complexity
was introduced by Cook in 1971
\cite{Cook-ProcACM-1971,Karp-CoCC-1972,Garey.Johnson-1979,Papadimitriou.Steiglitz-1998},
which distinguishes computation problems according to how the
computing time scales with the size of the problem.
Class P (\textsl{polynomial})  problems can be solved in polynomial time,
namely the computing time $t$ is bounded by a polynomial function
 of the problem's number $N$ of variables, $t = O(N^c)$ with $c$ being a
 finite constant.
Class NP (\textsl{nondeterministic polynomial}) problems, however, may need an exponentially increasing time ($t\sim e^{c N}$) to solve in the
worst case.  The most difficult problems in the class NP are
referred to as NP-complete problems, which are
problems that can be mutually converted into
each other by a polynomial algorithm.
If one can solve \emph{all the instances} of one
 NP-complete problem in polynomial time, she or he
 can simultaneously solve all the NP-complete problems in polynomial time.
The existence or not of a polynomial-time algorithm for NP-complete
problems is the famous and basic  \emph{P$=?$NP} problem of computation complexity.

Whether a problem belongs to the NP-complete class is judged by the
worst-case computation difficulty. However, a \emph{typical} problem
instance of a NP-complete problem might actually be very easy to solve.
In the last twenty years, typical-case computation
complexity has been intensively studied as an emerging
interdisciplinary research topic of
mathematics, theoretical computer science and statistical physics
\cite{Cheeseman-Kanefsky-Taylor-1991}.
Statistical physics concepts and methods have played a very
significant role in understanding typical-case computation complexity
\cite{
Mezard.Parisi.Virasoro-SGTaB-1987,
Nishimori-SGIP-2001,
Mezard.Montanari-2009,
Talagrand-SpinGlasses-2003,
Hartmann.Weigt-2005}.
Using the replica method
\cite{
Mezard.Parisi.Virasoro-SGTaB-1987,
Nishimori-SGIP-2001} and
the cavity method
\cite{
Mezard.Parisi-EPJB-2001,
Mezard.Parisi-JStatPhys-2003}
of spin glass mean field theory,
many interesting and fundamental
 optimization problems have been investigated in
the statistical physics community, such as
the travelling salesman problem
\cite{
Mezard.Parisi-JPhysique-1986},
the $K$-satisfiability problem
\cite{
Monasson.Zecchina-PRE-1997,
Mezard.Parisi.Zecchina-Science-2002,
Mezard.Zecchina-PRE-2002,
Montatari.RicciTersenghi.Semerjian-JStatMech-2008,
Krzakala.etal-PNAS-2007},
the exclusive-or-satisfiability (XOR-SAT) problem
\cite{
Franz.Leone.RicciTersenghi.Zecchina-PRL-2001,
Cocco.Dubois.Mandler.Monasson-PRL-2003,
Mezard.RicciTersenghi.Zecchina-JStatPhys-2003},
the vertex cover problem (or independent set) problem and
the hitting set problem
\cite{
Weigt.Hartmann-PRL-2000,
Weigt.Hartmann-PRL-2001,
Weigt.Hartmann-PRE-2001,
Hartmann.Weigt-JPhyA-2003,
Zhou-EPJB-2003,
Weigt.Zhou-PRE-2006,
Zhou.Zhou-PRE-2009,
Zhang.Zeng.Zhou-PRE-2009,
DallAsta.Pin.Ramezanpour-PRE-2009,
Mezard.Tarzia-PRE-2007,
Mezard.Tarzia.Toninelli-JPhysC-2008},
the graph coloring problem
\cite{
Mourik.Saad-PRE-2002,
Mulet.etal-PRL-2002,
Zdeborova.Krzakala-PRE-2007},
the maximal matching problem
\cite{
Karp.Sipser-IEEFoCS-1981,
Zhou.OuYang-arxiv-2003,
Zdeborova.Mezard-JStatPhys-2006},
and the feedback vertex set problem
\cite{Zhou-EurPhysJB-2013}.
In this review we take  a single prototypic
combinatorial optimization problem, the vertex cover problem,
as an example to introduce the ideas behind the 
statistical physical methods and
algorithms. Aimed at a reader outside the spin glass
research field, we do not discuss the technical details but
focus mainly on the intuitive physical picture behind the
message-passing equations. Hopefully a motivated
reader will easily grasp the essential ingredients of
the mean field approaches and further adjust the
methods to other optimization problems.

The layout of the paper is as follows.
Section~2 introduces the vertex cover problem.
Section~3 summarizes some mathematical results on the
minimal vertex cover problem.
Section~4 briefly mentions some local search algorithms.
We introduce in Sec.~5 the concept of long range frustration
and link it to computational difficulty. We then
introduce a spin glass model in Sec.~6 and discuss two
message-passing algorithms.
Some additional discussions are made in
Sec.~7.

\section{The vertex cover problem}
 \label{sec:vc}

\begin{figure}
\begin{center}
\includegraphics[width=0.5\textwidth]{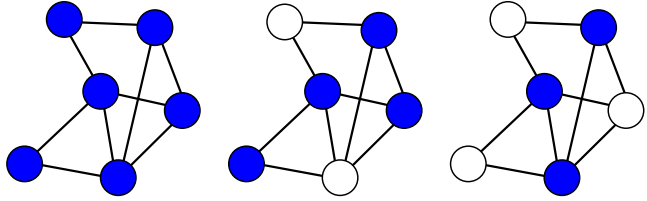}
\end{center}
\caption{
\label{fig:VCexamples}
Three vertex covers for a small graph of $N=6$ vertices and
$M=9$ edges. A vertex is represented by a filled circle if it belongs
to the VC, otherwise it is represented by an open circle.
}
\end{figure}

A graph $G=(V, E)$ is composed of a set $V$ of vertices and a
set $E$ of edges. There are $N$ vertices in the graph, therefore
$V = {1, 2, ..., N}$.
Each edge  connects between two different
vertices; for example, $(i, j)$ denotes an edge between
vertex $i$ and vertex $j$.
Here we consider sparse graphs such that the number of edges in graph
$G$ is of the same order as the number of vertices.
A vertex cover (VC) $V_{vc}$ of graph $G$ is a subset of vertices of the
graph which contains at least one incident vertices of every edge in
the set $E$. For example, if edge $(i, j) \in E$, then either
$i\in V_{vc}$ or $j\in V_{vc}$ or both. Figure~\ref{fig:VCexamples}
shows three VCs for a small graph.

With respective to a given vertex cover $V_{vc}$, a vertex $i$
is referred to as being covered if it belongs to $V_{vc}$,
otherwise the vertex is referred to as being uncovered.

The vertex cover problem is one of the first $21$ problems shown
to be NP-complete, it has fundamental importance in the
field of computation complexity \cite{Karp-CoCC-1972}.
This problem also have wide practical applictions,
for example internet traffic monitoring
\cite{Breitbart.etal-INFOCOM-2001},
prevention of denial-of-service attacks
\cite{Park.Lee-SIGCOM-2001},
immunization strategies in networks
\cite{GomezGardenes.Echenique.Moreno-EPJB-2006},
and network source location problem \cite{Huang.Raymond.Wong-2012}.

The vertex cover problem can be expressed either as a decision
problem or as an optimization problem.
As a decision problem, we ask whether there exists a vertex cover
$V_{vc}$ with cardinality $|V_{vc}|$ less then a certain given value.
As an optimization problem, we need to construct a vertex cover
whose cardinality is the global minimum over all possible
VCs for a given graph $G$.
In this review, we focus on the optimization problem for random
graphs. In our following discussions, we refer to the relative size (with respect to the vertex number $N$)
of a vertex cover $V_{vc}$ as its energy density and denote it
by $x$, namely
\begin{equation}
x \equiv \frac{|V_{vc}|}{N} \; .
\end{equation}
The global minimal energy density for a given graph is denoted as
$x_0$. If the cardinality of a VC for a given graph is the global minimum among all the VCs, it is referred to as an optimal VC.

\section{Mathematical Bounds and Asymptotics}
 \label{sec:bounds}

Here we list some established results on
rigorous bounds and asymptotic behaviors concerning the
global minimum of VCs.

On a general graph $G$, Harant's upper bound \cite{Harant-DiscrMath-1998}
on the minimal energy density $x_0$, obtained by
generalizing the earlier work of references
\cite{
Caro-TechRep-1979,
Wei-BellLab-1981},
is expressed as
\begin{equation}
x_{c}(G)
 \le
 1 -
 \frac {1}{N} \frac {(\sum _{i \in V} \frac {1}{d_i + 1})^2}
 {\sum _{i \in V} \frac{1}{d_i + 1}
 - \sum _{(i, j) \in E} \frac{(d_i - d_j)^2}{(d_i + 1)(d_j + 1)}} \; ,
\end{equation}
where $d_i$ is the degree (the number of attached edges)
of vertex $i$.
More refined upper bounds of $x_0$, not in the form of explicit expressions,
can be found in \cite{
Harant-DiscAppMath-2011,
Harant-Discu.Math.GraphTheory-2006,
Angel.Campigotto.Laforest-DiscAppMath-2013}.

A random graph of mean vertex degree $c$ is obtained by setting up $M=
(c/2) N$ edges completely at random starting from an empty graph of $N$
vertices
\cite{
Bollabas-RandomGraphs-2001,
Erdos.Renyi-Hungary-1960}.
For such random graphs, the work of Gazmuri
\cite{Gazmuri-Network-1984} predicts that the minimal energy density
$x_0$ almost surely are bounded by
$x_l < x_0  < 1 - \ln c / c$,
where $x_l$ is the root of
\begin{equation}
 x \ln x+ (1 - x ) \ln (1 - x )
 + (c / 2) (1 - x)^2 = 0 \; .
\end{equation}
These bounds are further improved by using the method of weighted
second moment \cite{Dani-Moore-2011}.
In the case of $c\rightarrow \infty$, Frieze has obtained the
following asymptotic expression for the minimal energy
density
\cite{Frieze-DiscrMath-1990}:
\begin{equation}
x_0
 = 1 - (2 / c) (\ln c - \ln \ln c + 1 - \ln 2) + o(1/c) \; .
\end{equation}

\section{Some heuristic local algorithms}
 \label{sec:heuristics}

An algorithm that is guaranteed to find a VC of global minimal
cardinality is branch-and-bound (see
Ref.~\cite{Hartmann.Weigt-JPhyA-2003} for a detailed
description). This algorithm performs an optimized search over
all the VCs of a given graph to determine the global minimum of
VC cardinality. Since the search space increases exponentially
with vertex number $N$, this algorithm works only for
small graphs.

There exist also many heuristic algorithms which construct
VCs based on some local rules.
One very simple heuristic algorithm is \emph{maximum degree decimation},
which recursively adds a vertex of the largest degree into the
vertex cover set and then reduces the graph by deleting
this vertex and its connected edges.
We can improve the performance of this greedy algorithm by combining
it with a \emph{leaf-removal} process \cite{Bauer.Golinelli-EPJB-2001}.
A vertex $i$ is considered as a leaf vertex if
this vertex is attached by only one edge, say $(i, j)$.
For such a leaf vertex $i$, it is always an optimal choice to add
the neighboring vertex $j$ instead of vertex $i$ into a VC.
The combined heuristic algorithm then works as follows:
As long as there is a leaf vertex,
add the neighboring vertex of this leaf vertex to the VC and  simplify the graph
$G$, otherwise add a vertex of the largest degree into the VC and
simplify $G$. The performance of such an algorithm on random
graphs is shown in Fig.~\ref{fig:greedyVC}. When the mean vertex
degree $c<e=2.718\cdots$ this algorithm has high probability
of constructing a VC of global minimal cardinality,
but for $c>e$ the energy density of the constructed VC is higher than the minimal value $x_0$.

The vertex cover problem can also be
solved by Monte Carlo optimization methods
\cite{
Newman.Barkema-MonteCarlo-1999,
Landau.Binder-MonteCarlo-2013}.
For example,
Ref.~\cite{Bartel.Hartmann-PRE-2004} adopts
the parallel tempering technique
\cite{
Marinari.Parisi-EPL-1992,
Hukushima.Nemoto-JPhysSocJpn-1996}
and obtains near-optimal VCs for relatively large
single random graphs.

\section{Long range frustrations}

Theoretical analysis revealed that the structure of a random
graph has a continuous phase transition at mean connectivity
$c=e$, characterized by the emergence of a core of
macroscopic size \cite{
Bauer.Golinelli-EPJB-2001,
Zhou-PRL-2005,
Zhou-PRL-2012,
Liu.Csoka.Zhou.Posfai-PRL-2012}.
A random graph has no core if its mean vertex degree $c<e$,
therefore the leaf-removal process can delete all the edges of
the graph and construct an optimal VC.

\begin{figure}
\begin{center}
\includegraphics[width=0.5\textwidth]{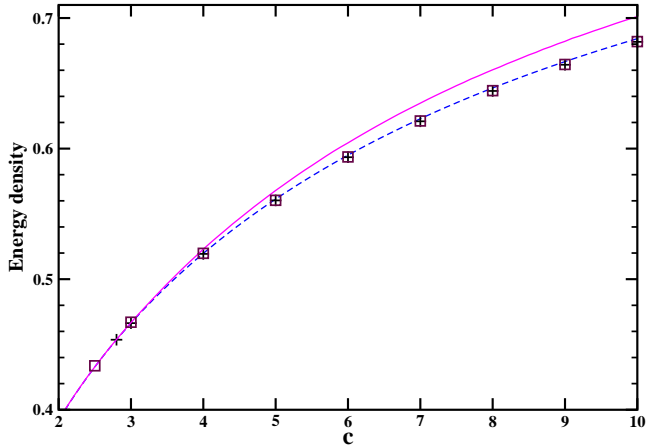}
\end{center}
\caption{ \label{fig:greedyVC}
  The solid line is the energy density $x(c)$ obtained by the
  hybrid algorithm of leaf-removal and maximum degree decimation
  on a single random graph of vertex number $N=10^6$ and
  mean vertex degree $c$.
  The dashed line is the energy density minimum $x_0(c)$
  predicted by the long range frustration theory at
  $N=\infty$. Each square symbol is the mean energy density obtained by
  the SPD algorithm on $16$ random graph instances of vertex
  number $N=10^5$,
  while each plus symbol shows the
  energy density minimum $x_0(c)$ predicted by the first-step replica-symmetry-breaking (1RSB)
  mean field theory at $N=\infty$.
}
\end{figure}

On the other hand,
the existence of a core at $c>e$ leads to very complicated
long range frustrations among the covering states of
different variables, making it impossible for a local
algorithm to find an optimal VC
\cite{Zhou-PRL-2005,Zhou-PRL-2012}. Let us focus on the
set $\Gamma_0$ of optimal VCs. If a vertex $i$ is always being
covered or always being uncovered in all the VCs of set 
$\Gamma_0$, then it is regarded as a frozen vertex with respect to $\Gamma_0$.
Otherwise vertex $i$ is an unfrozen vertex with respect to
$\Gamma_0$, meaning that it is being covered in some (but not
all) of the VCs of set $\Gamma_0$.
To explain the picture of long range frustrations, let us
randomly pick up two of such unfrozen vertices, say $j$ and $k$.
These two vertices might be far apart in terms of
shortest-distance path length.
There are four possible joint states for these two vertices. If
all these four joint states can be observed in at least one
VC of set $\Gamma_0$, then vertices $j$ and $k$ are regarded as
being unfrustrated, and fixing one vertex to the covered or
uncovered state will not affect the other vertex. However, if
at least one joint state (say vertex $j$ being covered and 
vertex $k$  being uncovered) of these two vertices is not
observed in any VC of set $\Gamma_0$, then these two vertices
are said to be (long-range) frustrated. In such a case,
fixing vertex $j$ to the covered state will cause vertex $k$
also to covered, even though $j$ and $k$ might be extremely
separated in the graph.

Such  complicated long range frustration effects can be
quantitatively considered by a mean field theory. For
a random graph of mean vertex degree $c$, the long range
frustration theory of 
Refs.~\cite{Zhou-PRL-2005,Zhou-PRL-2012} predicts
the minimal energy density $x_0$ to be
\begin{equation}
\label{eq:mvcelrf}
x_0 = 1 -\frac{1}{c} \int\limits_{0}^{c} r_0 ( \tilde{c}) {\rm d} \tilde{c} \;  .
\end{equation}
In this expression,
$r_0$ is the fraction of vertices that are not belonging to
any optimal VC, whose value is determined by solving
the following three equations involving $r_0$ and two
other quantities $r_*$ and $R$ (for details, see
\cite{Zhou-PRL-2005,Zhou-PRL-2012,Zhou-SGMP-2013}):
\begin{eqnarray}
r_0 & = &
2 e^{ -c r_0- c r_* R/2 } - e^{ -c r_0 - c r_* R } \; ,
\label{eq:r0} \\
r_* & = &
(2 c r_0 + c r_* R) e^{-c r_0 - c r_* R/2}
- \bigl(c r_0 + c r_* R+ (c r_* R)^2 /4 \bigr) e^{ -c r_0 - c r_* R } \; ,
\label{eq:rstar} \\
R & = &
\frac{c r_0^2}{r_*} \Bigl( 1 - \frac{1}{r_0} e^{- c r_0 -
c r_* R} \Bigr) \; .
\label{eq:lrfR}
\end{eqnarray}

The predicted values of $x_0(c)$ by this long range frustration
theory are in good agreement with the empirical
results obtained by the survey propagation-guided decimation
(SPD) algorithm \cite{Weigt.Zhou-PRE-2006} (see Section 6)
and with the theoretical results obtained by the
first-step replica-symmetry-breaking (1RSB)
mean field theory \cite{Zhou-EPJB-2003,Zhou-SGMP-2013},
see Fig.~\ref{fig:greedyVC}.
This indicates that the physical
picture behind the long range frustration theory is one of the
main reasons for the difficulty of obtaining
optimal VCs for random graphs with mean vertex degree $c>e$.
An extension of the long range frustration theory was made in
Ref.~\cite{Wei.Zhang.Guo.Zheng-PRE-2012}, which discussed the
backbone structure of optimal VCs.

\section{Message-passing algorithms}
 \label{sec:mpa}

For random graphs with mean vertex degree $c>e$,
because of the existence of long range frustrations among the
vertices, the minimal vertex cover problem is in a spin glass
phase. In such a phase, the optimal VCs are distributed into
many clusters. Each VC cluster contains a number of highly
similar VCs, while the VCs of different clusters are much
less similar with each other. Besides optimal VCs,
the system also have an enormous number of local minimal VCs,
which also form many clusters. The number of local minimal
VCs exponentially exceeds that of optimal VCs. Therefore
a local search algorithm will be trapped into one of the
local minimal VCs with certainty.

In the last twenty years,
the mean field theoretical methods of spin glasses, namely
the replica method
\cite{
Mezard.Parisi.Virasoro-SGTaB-1987,
Nishimori-SGIP-2001,Talagrand-SpinGlasses-2003, Binder.Young-RMP-1986}
and the cavity method
\cite{Mezard.Montanari-2009,Mezard.Parisi-EPJB-2001},
have been
applied on the vertex cover problem
\cite{Weigt.Hartmann-PRL-2000,
Weigt.Hartmann-PRE-2001,
Zhou-EPJB-2003,
Weigt.Zhou-PRE-2006,
Zhou-SGMP-2013} to describe its complex energy landscape.
From the physical point of view, the cavity method is based on the
Bethe-Peierls approximation of statistical physics
\cite{
Bethe-ProcRSoc-1935,
Peierls-ProRSoc-1936,
Peierls-ProcCambPhilSoc-1936}
and the physical picture that the configuration space can be regarded as
a collection of macroscopic states (each macroscopic state itself
contains a set of microscopic configurations). The cavity method
can also be understood from a more mathematical 
framework of partition function expansion
\cite{Xiao.Zhou-2011,Zhou.Wang-2012}.
This method  is
particularly convenient for investigating single problem
instances. Here we will describe two message-passing
algorithms inspired by the cavity method, namely
belief propagation and survey propagation.

\subsection{Spin glass model}

Since each vertex $i$ has two candidate covering states
$s_i=0$ (uncovered) and $s_i=1$ (covered), the total number of
microscopic configurations is $2^N$.
We introduce the following partition function $Z(\beta)$ for
the vertex cover problem
\cite{Weigt.Zhou-PRE-2006,Zhou-SGMP-2013}:
\begin{equation}
\label{eq:Zbeta}
Z(\beta)
 = \sum\limits_{\underline{s}}
 \prod\limits_{i = 1}^{N} e^{- \beta s_i}
 \prod\limits_{(j, k) \in G}
 \bigl[1 - (1 - s_j) (1 - s_k)\bigr] \; ,
\end{equation}
where the summation is over all the $2^N$ microscopic
configurations $\underline{s} \equiv \{s_1, s_2, \ldots, s_N\}$.
The edge product term of Eq.~(\ref{eq:Zbeta}) guarantees that $s_j+s_k \geq 1$ for each edge $(j, k)$ of the graph (i.e.,
at least one of the two vertices $j$ and $k$ is in the
covered state), otherwise the microscopic configuration
$\underline{s}$ have no contribution to $Z(\beta)$.
Therefore only VCs contribute to $Z(\beta)$.
The positive reweighting parameter $\beta$ emphasizes VCs of
smaller cardinality. In the limit of $\beta \rightarrow \infty$,
the partition function $Z(\beta)$ is contributed exclusively
by the optimal VCs.

\subsection{Belief Propagation}

Let us denote by $p_i^{(0)}$ the marginal probability that a
vertex $i$ is in the uncovering state $s_i=0$. Assuming the covering states of
 vertex $i$'s neighboring vertices are independent of
 each other \emph{in the absence of $i$} 
 (i.e., the Bethe-Peierls approximation),  we obtain the following expression for $p_i^{(0)}$:
\begin{equation}
\label{eq:pi}
p_{i}^{(0)}
 = \frac{
 \prod_{j \in \partial i} (1- p_{j \rightarrow i}^{(0)} )
 }
 {e^{- \beta} + 
 \prod_{j \in \partial i}
  (1- p_{j \rightarrow i}^{(0)})}
 \; ,
\end{equation}
where $\partial i$ denotes the set of neighboring vertices of vertex $i$, and
$p_{j\rightarrow i}^{(0)}$ is the probability of 
vertex $j$ being in the
uncovering state $s_j=0$ in the absence of the edge $(i, j)$. To understand
the above expression, we notice that, when the covering state of
vertex $i$ is $s_i=0$, all its neighboring vertices $j$ must be in the covering state $s_j=1$.
Such a requirement leads to the product term of the denominator
(and also that of the numerator). On the other hand, if
vertex $i$ is in the covered state, the constraints on
all the edges attached to $i$ are simultaneously satisfied, but
the VC cardinality increases by $1$, leading to a Boltzmann
factor $e^{-\beta}$ in the denominator of the above
expression.

Under the same Bethe-Peierls approximation we can write down the equation for
$p_{j\rightarrow i}^{(0)}$ as
\begin{equation}
\label{eq:bp}
p_{j \rightarrow i}^{(0)}
 = \frac {\prod_{k \in \partial j \backslash i}
 (1- p_{k \rightarrow j}^{(0)})}{e^{- \beta} + \prod_{k \in \partial j \backslash i}
 (1- p_{k \rightarrow j}^{(0)})} \; ,
\end{equation}
where $\partial k\backslash i$ denotes the set of neighboring vertices of
vertex $k$ (excluding vertex $i$). Equation~(\ref{eq:bp}) is referred to as
the belief propagation (BP) equation for the vertex cover problem. This
equation can also be derived from the framework of partition function
expansion \cite{Xiao.Zhou-2011,Zhou.Wang-2012}.  For a graph with $M$ edges there
are $2 M$ such equations.
We can try to solve this set of equations by numerical iteration. If a
fixed point can be reached by this iteration process, the energy density  $x$
(i.e., the mean fraction of vertices in the
covered state) is then evaluated as
\begin{equation}
x =  1- \frac{1}{N} \sum\limits_{i=1}^{N} p_i^{(0)} \; .
\end{equation}
The entropy density of VCs at the energy density $x$ can also
be computed \cite{Zhou.Zhou-PRE-2009}.

Based on Eqs.~(\ref{eq:pi}) and (\ref{eq:bp}), we have implemented a
simple belief propagation-guided decimation (BPD) algorithm as follows.
At a given value of $\beta$, we iterate the BP equation (\ref{eq:bp}) a
number of steps on a given graph $G$, and then compute the marginal probabilities $p_i^{(0)}$ for
all the vertices. Then a small fraction of vertices $i$
with  the smallest values of $p_i^{(0)}$ are added to an VC and deleted from the graph
$G$. We then simplify the graph $G$ and, if the simplified $G$ still contains
some edges, we repeat the above mentioned iteration-fixation
process. Figure~\ref{fig:bpsp} demonstrates that the performance of this BPD
algorithm is very good on random graphs. At each mean vertex degree $c$,
the energy density of the constructed VC by BPD is very close to the
energy density minimum $x_0(c)$ predicted by the 1RSB 
mean field theory.

\subsection{Survey Propagation}

Although the BPD algorithm seems to work excellently on single random
graph instances, the iteration of the BP equation (\ref{eq:bp}) actually
can not converge to a fixed point when the reweighting parameter $\beta$
is sufficiently large \cite{Zhang.Zeng.Zhou-PRE-2009}. The reason behind
this non-convergence is the breaking of ergodicity. When $\beta$ is sufficiently
large, we need to extend the Bethe-Peierls approximation to include
the possibility of the existence of many macroscopic states.

\begin{figure}
\begin{center}
\includegraphics[width=0.49\textwidth]{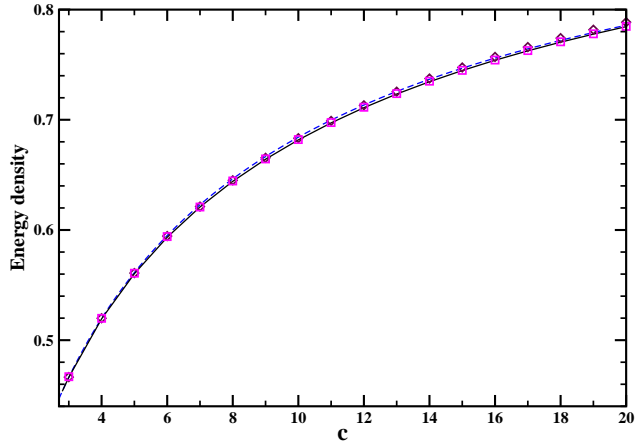}
\end{center}
\caption{
\label{fig:bpsp}
The solid line and the dashe line are, respectively,
the curve of energy density minimum $x_0(c)$
predicted by the 1RSB mean field theory and the
long range frustration theory for random graphs of mean
vertex degree $c$ and $N=\infty$.
Dimond symbols and square symbols are, respectively, the
VC energy densities reached by the
 BPD algorithm ($\beta = 10$) and the SPD algorithm
 ($y=3.05$) on a single
 random graph instance of $N=10^5$ vertices.
}
\end{figure}

As we are interested in the optimal VCs, we
focus on the limiting case of $\beta=\infty$. Then each
macroscopic state $\alpha$ of the configuration space is
characterized by a minimal energy density $x^{(\alpha)}$.
We can define a partition function at the level of
macroscopic states as
\begin{equation}
\Xi(y) = \sum\limits_{\alpha} e^{-y N x^{(\alpha)}} \; ,
\end{equation}
where the summation is over all the macroscopic states, and
$y$ is a reweighting parameter at the macroscopic states level.
A larger value of $y$ favors macroscopic states of smaller
minimal energies.

At $\beta = \infty$, the relevant microscopic configurations
must all be energy local or global minimal points of the
energy landscape. A macroscopic state is then
composed of a set of minimal energy configurations of the
same energy. In such a macroscopic state $\alpha$
we can describe the covering state of each vertex $i$ in
a coarse-grained way: (1) if the covering state $s_i = 1$ in all
the minimal energy configurations of $\alpha$, then $i$ is
said to be frozen to the covered state, with coarse-grained
state $S_i = 1$; (2) if $s_i=0$ in all the minimal energy
configurations of $\alpha$, then $i$ is said to be frozen
to the uncovered state, with coarse-grained state $S_i=0$;
(3) in the remaining cases, $s_i=0$ in some (but not all) the
minimal energy configurations and $s_i=1$ in the
remaining configurations of $\alpha$, and then we regard $i$
as being unfrozen, with coarse-grained state $S_i= *$.

Let us denote by $\pi_i^{(0)}$ the probability that vertex $i$
is in the coarse-grained state $S_i=0$. If we assume that
the coarse-grained states of the neighboring vertices of
$i$ are all independent \emph{in the absence of $i$} (this is
the Bethe-Peierls approximation at the level of
coarse-grained states), the following expression for
$\pi_i^{(0)}$ can be written down
\cite{Zhou-EPJB-2003,Weigt.Zhou-PRE-2006}:
\begin{equation}
\label{eq:cpi}
\pi_i^{(0)} =
\frac{\prod_{j \in \partial i}
(1 - \pi_{j \rightarrow i}^{(0)})}
{
e^{-y} + (1 - e^{-y})
\prod_{j \in \partial i}
(1 - \pi_{j \rightarrow i}^{(0)}) } \; ,
\end{equation}
where $\pi_{j\rightarrow i}^{(0)}$ is the probability that
the neighboring vertex $j$ is in the coarse-grained
covering state $S_j=0$ in the absence of vertex $i$.
The above expression has a clear intuitive interpretation.
If all the neighbors $j$ of the central vertex $i$ are not
in the coarse-grained covering state $S_j=0$ before vertex
$i$ is added to the graph, then there should exist at least
one macroscopic state in which all these neighboring vertices
are in the covered state. When $i$ is added to the
graph, then its covering state should be set to $s_i=0$ to
decrease energy. In all the other cases, the addition of
vertex $i$ will cause an increase of the minimal energy by
$1$, which explains the Boltzmann factor $e^{-y}$ in the above
equation.

Under similar considerations, we have the following equation
for the probability $\pi_{j\rightarrow i}^{(0)}$:
\begin{equation}
\label{eq:sp}
\pi_{j\rightarrow i}^{(0)} =
\frac{\prod_{k \in \partial j\backslash i}
(1 - \pi_{k \rightarrow j}^{(0)})}
{
e^{-y} + (1 - e^{-y})
\prod_{k \in \partial j\backslash i}
(1 - \pi_{k \rightarrow j}^{(0)}) } \; .
\end{equation}
This equation is referred to as the survey propagation (SP)
equation for the vertex cover problem \cite{Mezard.Parisi-EPJB-2001,
Mezard.Parisi.Zecchina-Science-2002,
Zhou-EPJB-2003,Weigt.Zhou-PRE-2006}.

The mean energy density and the entropy density $\Sigma$
at the level of macroscopic states both can be expressed as
functions of the probabilities $\{\pi_{i\rightarrow j}^{(0)},
\pi_{j\rightarrow i}^{(0)}: (i,j)\in G\}$. Such a
theoretical procedure is referred to as the first-step
replica-symmetry-breaking (1RSB) mean field procedure. It can
be justified again through the framework of partition function
expansion \cite{Xiao.Zhou-2011,Zhou.Wang-2012}.
In the numerical calculations, the reweighting
parameter $y$ is set to be the largest value such that
the entropy density $\Sigma$ is non-negative. The energy
density at this specific $y$ is then regarded as the
global minimum energy density. We shown in Fig.~\ref{fig:greedyVC} and Fig.~\ref{fig:bpsp} the ensemble-averaged minimum energy density obtained in such a way
as a function of mean vertex degree $c$. These figures show that
the 1RSB predictions are in close agreement with the predictions
of the long range frustration theory and with the empirical
results obtained by the
message-passing algorithms.

For single graph instances, similar to the BPD algorithm, we
can use the information obtained by the Eqs.~(\ref{eq:cpi}) and (\ref{eq:sp}) to construct near-optimal VCs. A survey
propagation-guided decimation (SPD) algorithms runs similarly
as the BPD algorithm: We iterate the SP equation
(\ref{eq:sp}) for a number of steps and then determine the
probability $\pi_i^{(0)}$ for each vertex $i$; then a small
fraction of vertices $i$ with the smallest values of
$\pi_i^{(0)}$ are declared as being covered; we then
simplify the graph $G$ and repeat the iteration-fixation process
as long as there are still edges in $G$.
The results of this SPD algorithm
are shown in Fig.~\ref{fig:bpsp} for random graphs. We find that
SPD very slightly outperforms BPD.

\section{Discussions}
 \label{sec:conclusion}

For purely random graphs, the 1RSB mean field theory and the
long range frustration theory both can give very good predictions
about the global minimal cardinality of vertex covers.
On the heuristic algorithms side, both the BPD
algorithm (inspired by the replica-symmetric mean field theory)
and the SPD algorithm (inspired by the 1RSB mean field theory)
are able to construct vertex covers whose cardinalities reach
the theoretically predicted values. These successes indicate
that the mean field theories of statistical physics can give
a good description about the statistical properties of the
random vertex cover problem. The good performance of the
BPD and SPD algorithms also means that near-optimal vertex
covers can be efficiently constructed for random graph instances.

The BPD and the SPD algorithms are comparable
in terms of implementation costs and computation time and
memory space.
As demonstrated in Fig.~\ref{fig:bpsp}, the SPD algorithm
slightly outperforms the BPD algorithm in terms of the
cardinality of constructed vertex covers.
Real-world instances of the vertex cover problem usually are
not random graphs but have certain structural properties. We
expect these two message-passing algorithms will also have
very good performances for such instances.

The minimal vertex cover problem is complementary to the
maximal independent set problem, which asks for the
construction of a smallest set of vertices such that any two
vertices of this set are not connected by an edge.
This later problem has significant applications
in game theory and microeconomics
\cite{
Bramoulle.Kranton-JEcoTheory-2007,
Galeotti.etal-NetworkGames-2009},
and glass transition
\cite{
Ritor.Sollich-AddPhys-2003}.
Since the independent set problem is equivalent to the
vertex cover problem, the methods and algorithms described
in this paper can be applied to this problem without much
modification. For example, Ref.~\cite{
DallAsta.Pin.Ramezanpour-PRE-2009}
has offered a detailed analysis of the
entropy density of independent sets.

The hitting set problem
\cite{
Mezard.Tarzia-PRE-2007}
is a natural extension of vertex cover problem.
It is a vertex cover problem defined on a hypergraph, in
which each edge may connect simultaneously with more than
two vertices.
This problem is also a NP-complete problem, and one of its
important applications is in group testing
\cite{
Mezard.Tarzia.Toninelli-JPhysC-2008}.
Another interesting extension
is the $k$-path vertex cover problem
\cite{Bresar.etal-DiscAppMath-2011},
which asks for the construction of a set of vertices intersecting
with every path of length $k \ge 2$ in a given graph (the case
$k=2$ is just the vertex cover problem).
The $k$-path vertex cover problem is originated from
the field of communication protocol \cite{Novotny-2010}.
It can be regarded as a special type of the hitting set problem.

\section{Acknowledgement}

J.-H. Zhao thanks Prof. Zhong-Can Ou-Yang for support.


\end{document}